\documentclass[aps,eqsecnum,preprint,floats,epsf,epsfig,nofootinbib]{revtex4}
\usepackage{graphicx}
\usepackage{epsfig}

\begin{document}
\def\be{\begin{eqnarray}}
\def\en{\end{eqnarray}}
\def\non{\nonumber}
\def\la{\langle}
\def\ra{\rangle}
\def\pp{{\prime\prime}}
\def\nc{N_c^{\rm eff}}
\def\vp{\varepsilon}
\def\hep{\hat{\varepsilon}}
\def\a{{\cal A}}
\def\B{{\cal B}}
\def\c{{\cal C}}
\def\d{{\cal D}}
\def\e{{\cal E}}
\def\p{{\cal P}}
\def\t{{\cal T}}
\def\up{\uparrow}
\def\dw{\downarrow}
\def\vma{{_{V-A}}}
\def\vpa{{_{V+A}}}
\def\smp{{_{S-P}}}
\def\spp{{_{S+P}}}
\def\J{{J/\psi}}
\def\ov{\overline}
\def\Lqcd{{\Lambda_{\rm QCD}}}
\def\pr{{Phys. Rev.}~}
\def\prl{{ Phys. Rev. Lett.}~}
\def\pl{{ Phys. Lett.}~}
\def\np{{ Nucl. Phys.}~}
\def\zp{{ Z. Phys.}~}
\def\lsim{ {\ \lower-1.2pt\vbox{\hbox{\rlap{$<$}\lower5pt\vbox{\hbox{$\sim$}
}}}\ } }
\def\gsim{ {\ \lower-1.2pt\vbox{\hbox{\rlap{$>$}\lower5pt\vbox{\hbox{$\sim$}
}}}\ } }

\font\el=cmbx10 scaled \magstep2{\obeylines\hfill February, 2004}

\vskip 1.5 cm

\centerline{\large\bf Covariant Light-Front Approach for $B\to
K^*\gamma$, $K_1\gamma$, $K^*_2\gamma$ Decays}
\bigskip
\centerline{\bf Hai-Yang Cheng and Chun-Khiang Chua}
\medskip
\centerline{Institute of Physics, Academia Sinica}
\centerline{Taipei, Taiwan 115, Republic of China}

\bigskip
\bigskip
\centerline{\bf Abstract}
\bigskip
\small
 Exclusive radiative $B$ decays, $B\to K^*\gamma$, $K_1(1270)\gamma$, $K_1(1400)\gamma$ and
$K_2^*(1430)\gamma$, are studied in the framework of a covariant
light-front quark model. The tensor form factor $T_1(q^2)$ at
$q^2=0$, which is relevant to the decay $B\to K^*\gamma$, is found
to be 0.24, substantially smaller than what expected from the
conventional light-front model or light-cone sum rules. Taking
into account the sizable next-to-leading order (NLO) corrections,
the calculated branching ratio of $B\to K^*\gamma$ agrees with
experiment, while most of the existing models predict too large
$B\to K^*\gamma$ compared to the data. The relative strength of
$B\to K_1(1270)\gamma$ and $B\to K_1(1400)\gamma$ rates is very
sensitive to the sign of the $K_1(1270)$--$K_1(1400)$ mixing
angle. Contrary to the other models in which $K_1(1270)\gamma$ and
$K_1(1400)\gamma$ rates are predicted to be comparable, it is
found that one of them is strongly suppressed owing to a large
cancellation between two different form factor terms. The
calculated branching ratio of $B\to K_2^*\gamma$ is in a good
agreement with experiment and this may imply the smallness of NLO
corrections to this radiative decay mode.

\eject
\section{Introduction}
Recently we have studied the decay constants and form factors of
the ground-state $s$-wave and low-lying $p$-wave mesons within a
covariant light-front approach \cite{CCH}. This formalism that
preserves the Lorentz covariance in the light-front framework has
been developed and applied successfully to describe various
properties of pseudoscalar and vector mesons \cite{Jaus99}. We
extended the covariant analysis of the light-front model in
\cite{Jaus99} to even-parity, $p$-wave mesons. With some explicit
examples, we have pointed out in \cite{CCH} that relativistic
effects could manifest in heavy-to-light transitions at maximum
recoil where the final-state meson could be highly relativistic
and hence there is no reason to expect that the non-relativistic
quark model is still applicable. For example, the $B\to a_1$ form
factor $V_0^{Ba_1}(0)$ is found to be 0.13 in the relativistic
light-front model \cite{CCH}, while it is as big as 1.01 in the
Isgur-Scora-Grinstein-Wise model \cite{ISGW}, a non-relativistic
version of the quark model.

In the present work we wish to apply the covariant light-front
approach to the exclusive radiative $B$ decays: $B\to K^*\gamma$,
$K_1\gamma$ and $K_2^*\gamma$ involving both $s$-wave and $p$-wave
mesons in the final states. They receive dominant contributions
from the short-distance electromagnetic penguin process $b\to
s\gamma$. \footnote{The electromagnetic penguin mechanism $b\to
s\gamma$ can also manifest in other two-body radiative decays of
bottom hadrons such as $B_s\to\phi\gamma$, $\Lambda_b\to
\Sigma^0\gamma,~\Lambda\gamma$, $\Xi_b\to\Xi\gamma$,
$\Omega_b\to\Omega\gamma$. These decays have been studied in
\cite{Cheng95}.} The radiative decay $B\to K^*\gamma$ was first
measured by CLEO \cite{CLEO93} a decade ago and more recently by
both $B$ factories: BaBar and Belle. The measured branching ratios
are
 \be
 \B(B^0\to K^{*0}\gamma) &=& \cases{ (4.55\pm0.70\pm0.34)\times
 10^{-5} & CLEO \cite{CLEO00} \cr (4.23\pm 0.40\pm0.22)\times
 10^{-5} & BaBar \cite{BaBar02} \cr (4.09\pm0.21\pm0.19)\times
 10^{-5} & Belle \cite{Belle03}, } \non \\
 \B(B^+\to K^{*+}\gamma) &=& \cases{ (3.76\pm0.86\pm0.28)\times
 10^{-5} & CLEO \cite{CLEO00} \cr (3.83\pm0.62\pm0.22)\times
 10^{-5} & BaBar \cite{BaBar02} \cr (4.40\pm0.33\pm0.24)\times
 10^{-5} & Belle \cite{Belle03}. }
 \en
Note that the Belle results are still preliminary. The average
branching ratios for the two modes are \cite{Nishida03}
 \be \label{eq:K*BRworld}
 \B(B^0\to K^{*0}\gamma) &=& (4.17\pm0.23)\times 10^{-5}, \non \\
  \B(B^+\to K^{*+}\gamma) &=& (4.18\pm0.32)\times 10^{-5}.
  \en
The decays $B^+\to K_1(1270)^+\gamma$ and $B^+\to
K_1(1400)^+\gamma$ have been searched by Belle \cite{Belle02} in
the $K^+\rho^0\gamma$ and $K^{*0}\pi^+\gamma$ final states,
respectively. Although a sizable signal was observed by Belle,
only upper limits were provided due to a lack of ability to
distinguish these resonances. As for $B\to K_2^*(1430)\gamma$,
CLEO \cite{CLEO00} has reported the first evidence with the
combined result
 \be
 \B(B\to K_2^*\gamma)= (1.66^{+0.59}_{-0.53}\pm0.13)\times
 10^{-5}.
 \en
The most recent Belle measurement \cite{Belle02} yields
 \be
 \B(B^0\to K_2^{*0}\gamma)=(1.3\pm0.5\pm0.1)\times 10^{-5},
 \en
while BaBar \cite{BaBarK2} obtained the preliminary results
 \be
  \B(B^0\to K_2^{*0}\gamma) &=& (1.22\pm0.25\pm0.11)\times
  10^{-5}, \non \\
 \B(B^+\to K_2^{*+}\gamma) &=& (1.44\pm0.40\pm0.13)\times 10^{-5}.
 \en

Theoretically, the nonfactorizable corrections to the decay $B\to
K^*\gamma$ have been studied in the QCD factorization approach
\cite{BBNS} to the next-to-leading order (NLO) in QCD and to the
leading order in the heavy quark limit
\cite{QCDfacBFS,QCDfacBB,Ali}. Using the light-cone sum rule
(LCSR) result of $0.38\pm0.06$~\cite{Ball} for the form factor
$T_1(0)$ to be defined below, it is found in
\cite{QCDfacBFS,QCDfacBB,Ali} that the NLO corrections yield an
enhancement of the $B\to K^*\gamma$ decay rate that can be as
large as 80\%. The enhancement is so large that the predicted
branching ratio disagrees with the observed one
(\ref{eq:K*BRworld}). We shall show in the present work that the
covariant light-front approach will lead to a form factor $T_1(0)$
much smaller than what expected from LCSR and the conventional
light-front model and yield a significantly improved agreement
with experiment.

For $B\to K_1\gamma$ decays, we will first use the covariant
light-front model to evaluate the tensor form factors in $B\to
K_{1A}$ and $B\to K_{1B}$ transitions, where $K_{1A}$ and $K_{1B}$
are the $^3P_1$ and $^1P_1$ states of $K_1$, respectively, and
then relate them to the physical $K_1$ states $K_1(1270)$ and
$K_1(1400)$. Since the $K_1(1270)$--$K_1(1400)$ mixing angle is
large, we shall see that one of the radiative decays, $B\to
K_1(1270)\gamma$ or $B\to K_1(1400)\gamma$, is strongly
suppressed, contrary to the other model predictions in which the
aforementioned two decay modes are comparable in their rates.

The paper is organized as follows. The formulism for the tensor
form factors evaluated in the covariant light-front model is
presented in Sec. II. The numerical results for form factors and
decay rates together with discussions are shown in Sec. III.
Conclusion is given in Sec. IV followed by an Appendix on the
heavy quark limit behavior of one of the tensor form factors.

\section{Formalism}

The matrix element for the $B\to K^*\gamma$ transition is given by
 \be
iM=\la \overline K^*(P^\pp,\vp^\pp)\gamma(q,\vp)|-iH_{\rm
eff}|\overline
 B(P^\prime)\ra,
 \en
where
 \be
 H_{\rm eff}&=&-\frac{G_{\rm F}}{\sqrt2} V^*_{ts} V_{tb} c_{11}
 Q_{11},
 \non\\
 Q_{11}&=&\frac{e}{8\pi^2} m_b \bar s \sigma_{\mu\nu} (1+\gamma_5)
 b F^{\mu\nu},
 \en
with $P^{\prime(\pp)}$ being the incoming (outgoing) momentum,
$\vp^{(\pp)}$ the polarization vector of $\gamma$ ($K^*$),
$V_{ij}$ the corresponding Cabbibo-Kobayashi-Maskawa (CKM) matrix
element and $c_{11}$ the Wilson coefficient. As will be seen
below, the inclusion of nonfactorizable corrections to $B\to
K^*\gamma$ will amount to replacing $c_{11}$ by the effective
parameter $a_{11}$ to be discussed below in Sec. III. In this work
we will calculate the $B\to K^*$ and  $B\to K_1, K^*_2$ transition
tensor form factors in the covariant light-front quark model and
obtain the corresponding radiative decay rates.

Tensor form factors for $B\to K^*,\,K_1,\,K^*_2$ transitions are
defined by
  \be
\la \overline K^*(P^\pp,\vp^\pp)|\bar si\sigma_{\mu\nu}q^\nu
(1+\gamma_5) b|\overline B(P^\prime)\ra
          &=& i \epsilon_{\mu\nu\lambda\rho} \vp^{\pp\nu *} P^\lambda q^\rho\, T_1(q^2)
\non\\
           &&  + (\vp^{\pp*}_\mu P\cdot q-P_\mu \vp^{\pp*}\cdot q)T_2(q^2)
\non\\     && +\vp^{\pp*}\cdot q\left(q_\mu-P_\mu\frac{q^2}{P\cdot
                 q}\right) T_3(q^2),
\non\\
\la \overline K_{1A,1B}(P^\pp,\vp^\pp)|\bar si\sigma_{\mu\nu}q^\nu
(1+\gamma_5) b|\overline B(P^\prime)\ra
          &=& i \epsilon_{\mu\nu\lambda\rho} \vp^{\pp\nu *} P^\lambda q^\rho\, Y_{A1,B1}(q^2)
\non\\
           &&  + (\vp^{\pp*}_\mu P\cdot q-P_\mu \vp^{\pp*}\cdot q)Y_{A2,B2}(q^2)
\non\\
           &&  +\vp^{\pp*}\cdot q \left(q_\mu-P_\mu\frac{q^2}{P\cdot q}\right) Y_{A3,B3}(q^2),
\non\\
\la \overline K_2^*(P^\pp,\vp^\pp)|\bar si\sigma_{\mu\nu}q^\nu
(1+\gamma_5) b|\overline B(P^\prime)\ra
          &=& -i \epsilon_{\mu\nu\lambda\rho} \vp^{\pp\nu\sigma *}P^\sigma P^\lambda q^\rho\, \frac
          {U_1(q^2)}{m_B}
\non\\
           &&  - (\vp^{\pp*}_{\mu\sigma} P\cdot q-P_\mu \vp^{\pp*}_{\sigma\rho}
           q^\rho)P^\sigma\frac{U_2(q^2)}{m_B}
\non\\
           &&  -\vp^{\pp*}_{\sigma\rho} P^\sigma q^\rho \left(q_\mu-P_\mu\frac{q^2}{P\cdot q}\right)
                \frac{U_3(q^2)}{m_B},
 \label{eq:ffs}
 \en
where $P=P^\prime+P^\pp$, $q=P^\prime-P^\pp$ and the convention
$\epsilon^{0123}=1$ is adopted. The physical strange axial-vector
$K_1(1270)$ and $K_1(1400)$ are the mixture of $K_{1A}$ and
$K_{1B}$ (we follow PDG \cite{PDG} to denote the $^3P_1$ and
$^1P_1$ states of $K_1$ by $K_{1A}$ and $K_{1B}$, respectively)
owing to the mass difference of the strange and non-strange light
quarks:
 \be \label{eq:K1mixing}
 K_1(1270)=K_{1A} \sin\theta+K_{1B}\cos\theta,
 \nonumber\\
 K_1(1400)=K_{1A} \cos\theta-K_{1B}\sin\theta.
 \en
The mixing angle $\theta$ will be discussed in the next section.
For the masses of $K_{1A}$ and $K_{1B}$, we follow
Ref.~\cite{Suzuki} to determine them from the mass relations
$m_{K_{1A}}^2=m_{K_1(1270)}^2+m_{K_1(1400)}^2-m_{K_{1B}}^2$ and
$2m_{K_{1B}}^2=m_{b_1(1232)}^2+m_{h_1(1380)}^2$.

\begin{figure}[t!]
\centerline{{\epsfxsize3 in \epsffile{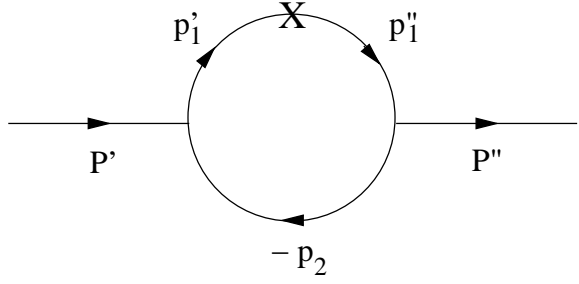}}}
%
\caption{Feynman diagrams for meson transition amplitudes, where
$P^{\prime(\pp)}$ is the incoming (outgoing) meson momentum,
$p^{\prime(\pp)}_1$ is the quark momentum, $p_2$ is the anti-quark
momentum and $X$ denotes the corresponding $\bar q^\pp
\sigma_{\mu\nu} (1+\gamma_5) q^\prime$ transition
vertex.}\label{fig:feyn}
\end{figure}

To begin with, we consider the transition amplitude given by the
one-loop diagram as shown in Fig.~\ref{fig:feyn}. 
We follow the approach of \cite{Jaus99} and use the same notation.
The incoming (outgoing) meson has the momentum
$P^{\prime(\pp)}=p_1^{\prime(\pp)}+p_2$, where $p_1^{\prime(\pp)}$
and $p_2$ are the momenta of the off-shell quark and antiquark,
respectively, with masses $m_1^{\prime(\pp)}$ and $m_2$. These
momenta can be expressed in terms of the internal variables $(x_i,
p_\bot^\prime)$,
 \be
 p_{1,2}^{\prime+}=x_{1,2} P^{\prime +},\qquad
 p^\prime_{1,2\bot}=x_{1,2} P^\prime_\bot\pm p^\prime_\bot,
 \en
with $x_1+x_2=1$. Note that we use $P^{\prime}=(P^{\prime -},
P^{\prime +}, P^\prime_\bot)$, where $P^{\prime\pm}=P^{\prime0}\pm
P^{\prime3}$, so that $P^{\prime 2}=P^{\prime +}P^{\prime
-}-P^{\prime 2}_\bot$.
In the covariant light-front approach, total four momentum is
conserved at each vertex where quarks and antiquarks are
off-shell. These differ from the conventional light-front approach
(see, for example, \cite{Jaus91,Cheng97}) where the plus and
transverse components of momentum are conserved, and quarks as
well as antiquarks are on-shell.
It is useful to define some internal quantities:
\begin{eqnarray} \label{eq:internalQ}
 M^{\prime2}_0
          &=&(e^\prime_1+e_2)^2=\frac{p^{\prime2}_\bot+m_1^{\prime2}}
                {x_1}+\frac{p^{\prime2}_{\bot}+m_2^2}{x_2},\quad\quad
                \widetilde M^\prime_0=\sqrt{M_0^{\prime2}-(m^\prime_1-m_2)^2},
 \nonumber\\
 e^{(\prime)}_i
          &=&\sqrt{m^{(\prime)2}_i+p^{\prime2}_\bot+p^{\prime2}_z},\quad\qquad
 p^\prime_z=\frac{x_2 M^\prime_0}{2}-\frac{m_2^2+p^{\prime2}_\bot}{2 x_2 M^\prime_0}.
 \end{eqnarray}
Here $M^{\prime2}_0$ can be interpreted as the kinetic invariant
mass squared of the incoming $q\bar q$ system, and $e_i$ the
energy of the quark $i$.

\begin{table}[b]
\caption{\label{tab:feyn} Feynman rules for the vertices
($i\Gamma^\prime_M$) of the incoming mesons-quark-antiquark, where
$p^\prime_1$ and $p_2$ are the quark and antiquark momenta,
respectively. Under the contour integrals to be discussed below,
$H^\prime_M$ and $W^\prime_M$ are reduced to $h^\prime_M$ and
$w^\prime_M$, respectively, whose expressions are given by
Eq.~(\ref{eq:h}). Note that for outgoing mesons, we shall use
$i(\gamma_0\Gamma^{\prime\dagger}_M\gamma_0)$ for the
corresponding vertices.}
\begin{tabular}{|c| c|}
\hline
 $M\,(^{2S+1}L_J) $
      &$i\Gamma^\prime_M$
      \\
      \hline
 pseudoscalar ($^1S_0$)
      &$H^\prime_P\gamma_5$
      \\
 vector ($^3S_1$)
      &$i H^\prime_V [\gamma_\mu-\frac{1}{W^\prime_V}(p^\prime_1-p_2)_\mu]$
      \\
 axial ($^3 P_1$)
      &$-i H^\prime_{^3\!A}[\gamma_\mu+\frac{1}{W^\prime_{^3\!A}}(p^\prime_1-p_2)_\mu]\gamma_5$
      \\
 axial ($^1 P_1$)
      &$-i H^\prime_{^1\!A} [\frac{1}{W^\prime_{^1\!A}}(p^\prime_1-p_2)_\mu]\gamma_5$
      \\
 tensor ($^3P_2$)
      &$i\frac{1}{2} H^\prime_T [\gamma_\mu-\frac{1}{W^\prime_V}(p^\prime_1-p_2)_\mu](p^\prime_1-p_2)_\nu$
      \\
\hline
\end{tabular}
\end{table}

It has been shown in \cite{CM69} that one can pass to the
light-front approach by integrating out the $p^-$ component of the
internal momentum in covariant Feynman momentum loop integrals.
We need Feynman rules for the meson-quark-antiquark vertices to
calculate the amplitudes depicted in Fig.~1. The Feynman rules for
vertices ($i\Gamma^\prime_M$) of ground-state $s$-wave mesons and
low-lying $p$-wave mesons are summarized in Table~\ref{tab:feyn}.
Note that we use $^3A$ and $^1A$ to denote $^3P_1$ and $^1P_1$
states, respectively. It is known that the integration of the
minus component of the internal momentum in Fig.~1 will force the
antiquark to be on its mass shell \cite{Jaus99}. The specific form
of the (phenomenological) covariant vertex functions for on-shell
quarks can be determined by comparing to the conventional
vertex functions~\cite{CCH}. 
%

We first consider the tensor form factors for $B\to K^*$
transition. We have
 \be
 {\cal B}_{\mu\nu}\vp^{\pp*\nu}
 \equiv\la \overline K^*(P^\pp,\vp^\pp)|\bar s\sigma_{\mu\lambda} q^\lambda
 (1+\gamma_5) b|\overline B(P^\prime)\ra
 =-i^3\frac{N_c}{(2\pi)^4}\int d^4 p^\prime_1
 \frac{H^\prime_P (i H^\pp_V)}{N_1^\prime N_1^\pp N_2} S_{R\mu\nu}\,\vp^{\pp*\nu},
 \label{eq:B}
 \en
where
 \be
S_{R\mu\nu} ={\rm
Tr}\left[\left(\gamma_\nu-\frac{1}{W^\pp_V}(p_1^\pp-p_2)_\nu\right)
                                 (\not \!p^\pp_1+m_1^\pp)
                                 \sigma_{\mu\lambda} q^\lambda (1+\gamma_5)
                                 (\not \!p^\prime_1+m_1^\prime)\gamma_5(-\not
                                 \!p_2+m_2)\right],
 \en
$N_1^\pp=p_1^{\pp2}-m_1^{\pp2}+i\epsilon$ and
$N_2=p_2^2-m_2^2+i\epsilon$. By using the identity
$2\sigma_{\mu\lambda}\gamma_5=i\epsilon_{\mu\lambda\rho\sigma}\sigma^{\rho\sigma}$,
the above trace $S_{R\mu\nu}$ can be further decomposed into
 \be
S_{R\mu\nu}=q^\lambda
S_{\nu\mu\lambda}+\frac{i}{2}q^\lambda\epsilon_{\mu\lambda\rho\sigma}
S_\nu^{\,\,\rho\sigma},
 \en
 where
  \be
S_{\nu\mu\lambda}&=&{\rm
Tr}\left[\left(\gamma_\nu-\frac{1}{W^\pp_V}(p_1^\pp-p_2)_\nu\right)
                                 (\not \!p^\pp_1+m_1^\pp)
                                 \sigma_{\mu\lambda} q^\lambda
                                 (\not \!p^\prime_1+m_1^\prime)\gamma_5(-\not
                                 \!p_2+m_2)\right]
 \non\\
           &=&-\epsilon_{\mu\nu\lambda\alpha}2[2(m^\prime_1
           m_2+m_1^\pp m_2-m^\prime_1 m_1^\pp)
           p_1^{\prime\alpha}+m_1^\prime m_1^\pp P^\alpha+(m_1^\prime
           m_1^\pp-2 m_1^\prime m_2) q^\alpha]
 \non\\
           &&-\frac{2}{W^\pp_V}(4 p^\prime_{1\nu}-3
           q_\nu-P_\nu)\epsilon_{\mu\lambda\alpha\beta}
           [(m_1^\prime+m_1^\pp)p^{\prime\alpha}_1 P^\beta
           +(m_1^\pp-m_1^\prime+2m_2) p_1^{\prime\alpha} q^\beta
 \non\\
           &&+m_1^\prime P^\alpha q^\beta].
   \label{eq:S}
 \en

As in \cite{Jaus99,CCH}, we work in the $q^+=0$ frame. For the
integral in Eq.~(\ref{eq:B}) we perform the $p_1^-$
integration~\cite{Jaus99}, which picks up the residue at $p_2=\hat
p_2$ and leads to
 \be
 N_1^{\prime(\pp)}
      &\to&\hat N_1^{\prime(\pp)}=x_1(M^{\prime(\pp)2}-M_0^{\prime(\pp)2}),
\non\\
 H^{\prime(\pp)}_M
      &\to& h^{\prime(\pp)}_M,
\non\\
 W^\pp_M
      &\to& w^\pp_M,
\non\\
\int \frac{d^4p_1^\prime}{N^\prime_1 N^\pp_1 N_2}H^\prime_P
H^\pp_V S
      &\to& -i \pi \int \frac{d x_2 d^2p^\prime_\bot}
                             {x_2\hat N^\prime_1
                             \hat N^\pp_1} h^\prime_P h^\pp_V \hat S,
 \label{eq:contourB}
 \en
where
 \be
 M^{\pp2}_0
          =\frac{p^{\pp2}_\bot+m_1^{\pp2}}
                {x_1}+\frac{p^{\pp2}_{\bot}+m_2^2}{x_2},
 \en
with $p^\pp_\bot=p^\prime_\bot-x_2\,q_\bot$. In this work the
explicit forms of $h^\prime_M$ and $w^\prime_M$ are given
by~\cite{CCH}
\begin{eqnarray} \label{eq:vertex}
 h^\prime_P&=&h^\prime_V
                  =(M^{\prime2}-M_0^{\prime2})\sqrt{\frac{x_1 x_2}{N_c}}
                    \frac{1}{\sqrt{2}\widetilde M^\prime_0}\varphi^\prime,
 \nonumber\\
 h^\prime_{^3\!A}
                  &=&(M^{\prime2}-M_0^{\prime2})\sqrt{\frac{x_1 x_2}{N_c}}
                    \frac{1}{\sqrt{2}\widetilde M^\prime_0}\frac{\widetilde
                     M^{\prime
                     2}_0}{2\sqrt{2}M^\prime_0}\varphi^\prime_p,
                        \nonumber\\
 h^\prime_{^1\!A}&=& h^\prime_T =(M^{2\prime}-M_0^{\prime 2})\sqrt{\frac{x_1
 x_2}{N_c}}\frac{1}{\sqrt{2}\widetilde M^\prime_0}\varphi'_p\, ,
 \non\\
 w^\prime_V&=&M^\prime_0+m^\prime_1+m_2,\quad
 w^\prime_{^3\!A}=\frac{\widetilde{M}'^2_0}{m^\prime_1-m_2},\quad
 w^\prime_{^1\!A}=2\,,
 \label{eq:h}
\end{eqnarray}
where $\varphi'$ and $\varphi'_p$ are the light-front momentum
distribution amplitudes for $s$-wave and $p$-wave mesons,
respectively. There are several popular phenomenological
light-front wave functions that have been employed to describe
various hadronic structures in the literature. In the present
work, we shall use the Gaussian-type wave function \cite{Gauss}
\begin{eqnarray} \label{eq:Gauss}
 \varphi^\prime
    &=&\varphi^\prime(x_2,p^\prime_\perp)
             =4 \left({\pi\over{\beta^{\prime2}}}\right)^{3\over{4}}
               \sqrt{{dp^\prime_z\over{dx_2}}}~{\rm exp}
               \left(-{p^{\prime2}_z+p^{\prime2}_\bot\over{2 \beta^{\prime2}}}\right),
\nonumber\\
 \varphi^\prime_p
    &=&\varphi^\prime_p(x_2,p^\prime_\perp)=\sqrt{2\over{\beta^{\prime2}}}~\varphi^\prime,\quad\qquad
         \frac{dp^\prime_z}{dx_2}=\frac{e^\prime_1 e_2}{x_1 x_2 M^\prime_0}.
 \label{eq:wavefn}
\end{eqnarray}
The parameter $\beta'$ is expected to be of order $\Lambda_{\rm
QCD}$.

In general, $\hat p^\prime_1$ can be expressed in terms of three
external vectors, $P^\prime$, $q$ and $\tilde\omega$
[$\tilde\omega$ being a lightlike vector with the expression
$\tilde\omega^\mu=(\tilde\omega^-,\tilde\omega^+,\tilde\omega_\bot)=(2,0,0_\bot)$].
In practice, for $\hat p_1^\prime $ under integration we use the
following rules~\cite{Jaus99}
 \be
\hat p^\prime_{1\mu}
       &\doteq& P_\mu A_1^{(1)}+q_\mu A_2^{(1)},
 \non\\
\hat p^\prime_{1\mu} \hat p^\prime_{1\nu}
       &\doteq& g_{\mu\nu} A_1^{(2)}+P_\mu P_\nu A_2^{(2)}+(P_\mu
                q_\nu+ q_\mu P_\nu) A^{(2)}_3+q_\mu q_\nu A^{(2)}_4,
 \non\\
\hat p^\prime_{1\mu} \hat p^\prime_{1\nu} \hat p^\prime_{1\alpha}
       &\doteq& (g_{\mu\nu} P_\alpha+g_{\mu\alpha} P_\nu+g_{\nu\alpha} P_\mu) A_1^{(3)}
               +(g_{\mu\nu} q_\alpha+g_{\mu\alpha} q_\nu+g_{\nu\alpha} q_\mu) A_2^{(3)}
 \non \\
       &&       +P_\mu P_\nu P_\alpha A_3^{(3)}
                +(P_\mu P_\nu q_\alpha+ P_\mu q_\nu P_\alpha+q_\mu P_\nu P_\alpha) A^{(3)}_4
 \non \\
       &&       +(q_\mu q_\nu P_\alpha+ q_\mu P_\nu q_\alpha+P_\mu q_\nu q_\alpha)
                 A^{(3)}_5
                +q_\mu q_\nu q_\alpha  A^{(3)}_6,
 \label{eq:p1B}
 \en
where the symbol $\doteq$ reminds us that the above equations are
true only after integration. In the above equation, $A^{(i)}_j$
are functions of $x_{1,2}$, $p^{\prime2}_\bot$,
$p^\prime_\bot\cdot q_\bot$ and $q^2$, and their explicit
expressions are given by~\cite{Jaus99}
 \be \label{eq:Aij}
 A^{(1)}_1&=&\frac{x_1}{2},
 \quad
 A^{(1)}_2=A^{(1)}_1-\frac{p^\prime_\bot\cdot q_\bot}{q^2},
\non\\
 A^{(2)}_1&=&-p^{\prime2}_\bot-\frac{(p^\prime_\bot\cdot q_\bot)^2}{q^2},
 \quad
 A^{(2)}_2=\big(A^{(1)}_1\big)^2,
 \quad
 A^{(2)}_3=A^{(1)}_1 A^{(1)}_2,
 \non\\
 A^{(2)}_4&=&\big(A^{(1)}_2\big)^2-\frac{1}{q^2}A^{(2)}_1,
 \quad
 A^{(3)}_1=A^{(1)}_1 A^{(2)}_1,
 \quad
 A^{(3)}_2=A^{(1)}_2 A^{(2)}_1,
 \\
 A^{(3)}_3&=&A^{(1)}_1 A^{(2)}_2,
 \quad
 A^{(3)}_4=A^{(1)}_2 A^{(2)}_2,
 \quad
 A^{(3)}_5=A^{(1)}_1 A^{(2)}_4,
  \non\\
 A^{(3)}_6&=&A^{(1)}_2 A^{(2)}_4-\frac{2}{q^2}A^{(1)}_2 A^{(2)}_1.
  \non
 \en
We do not show the spurious contributions in Eq.~(\ref{eq:p1B})
since they are numerically vanishing~\cite{Jaus99,Jaus03,CCH}. For
the integration in Eq.~(\ref{eq:B}) we need only the first two
rules in (\ref{eq:p1B}), while the third one will be used in the
calculation of the $B\to K_2^*$ transition form factors. In
general, there are additional rules involving $N_2$
in~\cite{Jaus99} and these may be identified as zero mode
contributions to form factors~(for a different approach of zero
mode contributions, see \cite{BCJ03}). As shown in
Eq.~(\ref{eq:S}), there is no $N_2$ term in the trace and hence no
zero mode contribution to the $B\to K^*$ form factors. As we shall
see, the above statement also holds for $B\to K_1$ and $B\to
K_2^*$ form factors.

By using Eqs.~(\ref{eq:B})--(\ref{eq:p1B}), one arrives at
 \be
  T_1(q^2)&=&\frac{N_c}{16\pi^3}\int dx_2 d^2 p^\prime_\bot
           \frac{2 h^\prime_P h^\pp_V}{x_2 \hat N^\prime_1 \hat N^\pp_1}
           \bigg\{m_1^\prime m_1^\pp+x_1 (m_1^\prime m_2+m_1^\pp m_2-m_1^\prime m_1^\pp)
           -\frac{2}{w^\pp_V}[(m_1^\prime+m_1^\pp)A^{(2)}_1]\bigg\},
 \non\\
 T_2(q^2)&=&T_1(q^2)+\frac{q^2}{P\cdot q} \frac{N_c}{16\pi^3}\int dx_2 d^2 p^\prime_\bot
           \frac{2 h^\prime_P h^\pp_V}{x_2 \hat N^\prime_1 \hat N^\pp_1}
           \bigg\{m_1^\prime m_1^\pp-2m_1^\prime m_2
 \non\\
           &&+2 A^{(1)}_2 (m_1^\prime m_2+m_1^\pp m_2-m_1^\prime m_1^\pp)
           -\frac{2}{w^\pp_V}[(m_1^\pp-m_1^\prime+2 m_2)A^{(2)}_1]\bigg\},
 \non\\
 T_3(q^2)&=&\frac{N_c}{16\pi^3}\int dx_2 d^2 p^\prime_\bot
           \frac{2 h^\prime_P h^\pp_V}{x_2 \hat N^\prime_1 \hat N^\pp_1}
           \bigg\{2m_1^\prime m_2-m_1^\prime m_1^\pp-2 A^{(1)}_2 (m_1^\prime m_2+m_1^\pp m_2-m_1^\prime m_1^\pp)
 \non\\
           &&+\frac{2}{w^\pp_V}\{(m_1^\pp-m_1^\prime+2
           m_2)[A^{(2)}_1+P\cdot q (A^{(2)}_2+A^{(2)}_3-A^{(1)}_1)
 \non\\
           &&+P\cdot q (m_1^\prime+m_1^\pp)
           (A^{(1)}_2-A^{(2)}_3-A^{(2)}_4)+P\cdot q\, m_1^\prime
           (A^{(1)}_1+A^{(1)}_2-1)
           ]\}\bigg\}.
 \label{eq:T}
 \en
In order to compare with the conventional light-front model
calculation for $T_1(0)$, which is relevant for $B\to K^*\gamma$
decay, we write
 \be \label{eq:T1}
 T_1(0) &=& {1\over 32\pi^2}\int dx\, d^2p'_\bot {\varphi^\pp(x,p'_\bot)\varphi'(x,p'_\bot)
 \over \sqrt{{\cal A}'^2+p'^2_\bot}\sqrt{{\cal A}^{\pp
 2}+p'^2_\bot}} \non \\
 &\times& \left\{x^2 m_bm_s+x(1-x)(m_bm_q+m_sm_q)+{p'^2_\bot\over
 \omega^\pp_V}x(m_b+m_s)\right\}.
 \en
where ${\cal A}'=m_bx+m_q(1-x)$ and ${\cal A}^\pp=m_sx+m_q(1-x)$,
$x=x_2$, and $m_q$ is the mass of the spectator quark in the $B$
meson. This is to be compared with the result
  \be \label{eq:T1ODonnell}
 T_1(0) &=& {1\over 32\pi^2}\int dx\, d^2p'_\bot {\varphi^\pp(x,p'_\bot)\varphi'(x,p'_\bot)
 \over \sqrt{{\cal A}'^2+p'^2_\bot}\sqrt{{\cal A}^{\pp
 2}+p'^2_\bot}}  \\
 &\times& \left\{x^2 m_bm_s+x(1-x)(m_bm_q+m_sm_q)+(1-x)[(1-x)m_q^2+p'^2_\bot]+{p'^2_\bot\over
 \omega^\pp_V}x(m_b+m_s)\right\} \non
 \en
obtained in \cite{ODonnell95}. It is clear that the terms
proportional to $(1-x)m_q^2+p'^2_\bot$ do not exist in our
expression for $T_1(0)$. This will affect the numerical result
significantly for the $B\to K^*\gamma$ rate as we shall discuss in
Sec. III. It is shown in Appendix that our result (\ref{eq:T1})
for $T_1(0)$ has the correct heavy quark limit behavior and hence
it is more trustworthy than (\ref{eq:T1ODonnell}).

The calculation for $B\to K_{1A,1B}$ transition form factors can
be done in a similar manner. In analogue to Eq.~(\ref{eq:B}), we
have
 \be
 {\cal B}^{^3A}_{\mu\nu}\vp^{\pp*\nu}
  &=&-i^3\frac{N_c}{(2\pi)^4}\int d^4 p^\prime_1
 \frac{H^\prime_P (-i H^\pp_{^3A})}{N_1^\prime N_1^\pp N_2} S^{^3A}_{R\mu\nu}\,\vp^{\pp*\nu},
 \non\\
 {\cal B}^{^1A}_{\mu\nu}\vp^{\pp*\nu}
  &=&-i^3\frac{N_c}{(2\pi)^4}\int d^4 p^\prime_1
 \frac{H^\prime_P (-i H^\pp_{^1A})}{N_1^\prime N_1^\pp N_2} S^{^1A}_{R\mu\nu}\,\vp^{\pp*\nu},
 \label{eq:Baxial}
 \en
where
 \be
S^{^3A}_{R\mu\nu} &=&{\rm
Tr}\left[\left(\gamma_\nu-\frac{1}{W^\pp_{^3A}}(p_1^\pp-p_2)_\nu\right)\gamma_5
                                 (\not \!p^\pp_1+m_1^\pp)
                                 \sigma_{\mu\lambda} q^\lambda (1+\gamma_5)
                                 (\not \!p^\prime_1+m_1^\prime)\gamma_5(-\not
                                 \!p_2+m_2)\right],
 \non\\
S^{^1A}_{R\mu\nu} &=&{\rm
Tr}\left[\left(-\frac{1}{W^\pp_{^1A}}(p_1^\pp-p_2)_\nu\right)\gamma_5
                                 (\not \!p^\pp_1+m_1^\pp)
                                 \sigma_{\mu\lambda} q^\lambda (1+\gamma_5)
                                 (\not \!p^\prime_1+m_1^\prime)\gamma_5(-\not
                                 \!p_2+m_2)\right].
 \en
It can be easily shown that $S^{^3A,^1A}_{R\mu\nu}=-S_{R\mu\nu}$
with $m_1^\pp$ and $W^\pp_V$ replaced by $-m_1^\pp$ and
$W^\pp_{^3A,^1A}$, respectively, while only the $1/W^\pp_{^1A}$
term is kept for the $S^{^1A}_R$ case. Consequently, we have, for
$i=1,2,3$,
 \be\label{eq:TA}
 Y_{Ai,Bi}(q^2)&=&T_i(q^2) \,\,\,{\rm with}\,\,\,
                         (m_1^\pp\to -m_1^\pp,\,h^\pp_V\to h^\pp_{^3\!A,^1\!A},\,w^\pp_V\to w^\pp_{^3\!A,^1\!A}),
 \en
where only the $1/W^\pp$ terms in $Y_{Bi}$ form factors are kept.
It should be cautious that the replacement of $m_1^\pp\to
-m_1^\pp$ should not be applied to $m_1^\pp$ in $w^\pp$ and
$h^\pp$. The above simple relation between $B\to K_1$ and $B\to
K^*$ transition tensor form factors is similar to that for vector
and axial form factors in $P\to A$ and $P\to V$ transitions
\cite{CCH}.

Finally we turn to the $B\to K^*_2$ transition given by
 \be
 {\cal B}^{T}_{\mu\nu\lambda}\vp^{\pp*\nu\lambda}
 \equiv
 \la K^*_2(P^\pp,\vp^\pp)|\bar s\sigma_{\mu\nu}(1+\gamma_5) q^\nu b|B(P^\prime)\ra
 =-i^3\frac{N_c}{(2\pi)^4}\int d^4 p^\prime_1
 \frac{H^\prime_P (i H^\pp_T)}{N_1^\prime N_1^\pp N_2} S^{PT}_{\mu\nu\lambda}\,\vp^{\pp*\nu\lambda},
 \label{eq:BT}
 \en
where
 \be
S^T_{R\mu\nu\lambda} =S_{R\mu\nu}(-q+p^\prime_1)_\lambda.
 \label{eq:ST}
 \en
The contribution from the $S_{\mu\nu}(-q)_\lambda$ part is
trivial, since $q_\lambda$ can be taken out from the integration,
which is already done in the $B\to K^*$ case. Contributions from
the $\hat S_{R\mu\nu}\hat p^\prime_{1\lambda}$ part can be worked
out by using Eq.~(\ref{eq:p1B}). Putting all these together leads
to
 \be
  U_1(q^2)&=&\frac{N_c}{16\pi^3}\int dx_2 d^2 p^\prime_\bot
           \frac{2M^\prime h^\prime_P h^\pp_V}{x_2 \hat N^\prime_1 \hat N^\pp_1}
           \bigg\{m_1^\prime m_1^\pp(1-A^{(1)}_1-A^{(1)}_2)
  \non\\
           &&+2 (m_1^\prime m_2+m_1^\pp m_2-m_1^\prime m_1^\pp)
              (A^{(1)}_1-A^{(2)}_2-A^{(2)}_3)
             -\frac{4}{w^\pp_V}\left[(m_1^\prime+m_1^\pp)(A^{(2)}_1-A^{(3)}_1-A^{(3)}_2)\right]\bigg\},
 \non\\
 U_2(q^2)&=&U_1(q^2)+\frac{q^2}{P\cdot q} \frac{N_c}{16\pi^3}\int dx_2 d^2 p^\prime_\bot
           \frac{2M^\prime h^\prime_P h^\pp_V}{x_2 \hat N^\prime_1 \hat N^\pp_1}
           \bigg\{(m_1^\prime m_1^\pp-2m_1^\prime
           m_2)(1-A^{(1)}_1-A^{(1)}_2)
 \non\\
           &&+2  (m_1^\prime m_2+m_1^\pp m_2-m_1^\prime m_1^\pp)
           (A^{(1)}_2-A^{(2)}_3-A^{(2)}_4)
 \non\\
           &&-\frac{4}{w^\pp_V}\left[(m_1^\pp-m_1^\prime+2 m_2)(A^{(2)}_1-A^{(3)}_1-A^{(3)}_2)\right]\bigg\},
 \non\\
 U_3(q^2)&=&\frac{N_c}{16\pi^3}\int dx_2 d^2 p^\prime_\bot
           \frac{2M^\prime h^\prime_P h^\pp_V}{x_2 \hat N^\prime_1 \hat N^\pp_1}
           \bigg\{(m_1^\prime m_1^\pp-2m_1^\prime m_2)(-1+A^{(1)}_1+A^{(1)}_2)
 \non\\
           &&+2 (m_1^\prime m_2+m_1^\pp m_2-m_1^\prime m_1^\pp)
           (-A^{(1)}_2+A^{(2)}_3+A^{(2)}_4)
 \non\\
           &&-\frac{2}{w^\pp_V}\Big[2(m_1^\pp-m_1^\prime+2
           m_2)(-A^{(2)}_1+A^{(3)}_1+A^{(3)}_2)
 \non\\
           &&+P\cdot q (m_1^\pp-m_1^\prime+2 m_2)(A^{(1)}_1-2A^{(2)}_2-2 A^{(2)}_3+A^{(3)}_3+2 A^{(3)}_4+A^{(3)}_5)
 \non\\
           &&+P\cdot q (m_1^\prime+m_1^\pp)
           (-A^{(1)}_2+2A^{(2)}_3+2A^{(2)}_4-A^{(3)}_4-2A^{(3)}_5-A^{(3)}_6)
 \non\\
           &&+P\cdot q \,m_1^\prime (1-2A^{(1)}_1-2A^{(1)}_2+A^{(2)}_2+2A^{(2)}_3+A^{(2)}_4)
           \Big]\bigg\}.
 \label{eq:U}
 \en

We are now ready to calculate the radiative decay rates. Before
proceeding, two remarks are in order: (i) At $q^2=0$ the form
factors obey the simple relations $T_2(0)=T_1(0)$,
$Y_{A2,B2}(0)=Y_{A1,B1}(0)$ and $U_2(0)=U_1(0)$. (ii)~Form factors
$T_3(0),\,Y_{3A,3B}(0),\,U_3(0)$ do not contribute to the
corresponding radiative decay rates. It is straightforward to
obtain
 \be
 {\mathcal B}(B\to K^*\gamma)&=&\tau_B\frac{G_{\rm F}^2 \alpha m_B^3
                                m_b^2}{32\pi^4}\left(1-\frac{m^2_{K^*}}{m_B^2}\right)^3|V_{tb}
                                 V^*_{ts} a_{11}\, T_1(0)|^2,
 \non\\
 {\mathcal B}(B\to K_{1A,1B}\,\gamma)&=&\tau_B\frac{G_{\rm F}^2 \alpha m_B^3
                                m_b^2}{32\pi^4}\left(1-\frac{m^2_{K_{1A,1B}}}{m_B^2}\right)^3|V_{tb}
                                 V^*_{ts} a_{11}\, Y_{A1,B1}(0)|^2,
 \non\\
 {\mathcal B}(B\to K^*_2\gamma)&=&\tau_B\frac{G_{\rm F}^2 \alpha m_B^5
                                m_b^2}{256\pi^4 m_{K^*_2}^2}\left(1-\frac{m^2_{K^*_2}}{m_B^2}\right)^5|V_{tb}
                                 V^*_{ts} a_{11}\, U_1(0)|^2,
 \label{eq:Br}
 \en
where $\tau_B$ is the $B$ lifetime. It has been realized recently
that non-factorizable strong interaction corrections (i.e. those
corrections not related to form factors, such as hard vertex and
hard spectator contributions) to $B\to K^*\gamma$ are calculable
in the heavy quark limit and amount to replacing the Wilson
coefficient $c_{11}$ by the effective parameter $a_{11}$. Such
corrections have been calculated in the QCD factorization
framework and in the large energy effective theory up to NLO in
$\alpha_s$ and to the leading power in $\Lambda_{\rm QCD}/m_B$ and
found to be quite sizable~\cite{QCDfacBFS,QCDfacBB,Ali}. We will
return back to this point later.

In the next section, we will give numerical results for form
factors, $T_i(q^2),\,Y_{Ai,Bi}(q^2),\,U_i(q^2)$, as well as $B\to
K^*\gamma,\,K_1\gamma,\,K^*_2\gamma$ decay rates.

\section{Numerical Results and Discussion}

To perform numerical calculations we need to first specific some
input parameters in the covariant light-front model. The input
parameters $m_q$ and $\beta$ in the Gaussian-type wave function
(\ref{eq:wavefn}) are shown in Table~\ref{tab:input}. The
constituent quark masses are close to those used in the
literature~\cite{Jaus96,Cheng97,Hwang02,Jaus99,CCH}. The input
parameters $\beta$'s are fixed by the decay constants whose
analytic expressions in the covariant light-front model are given
in \cite{CCH}. We use $f_B=180$~MeV and $f_{K^*}=230$~MeV to fix
$\beta_{B}$ and $\beta_{K^*}$, respectively. For $p$-wave strange
mesons, we take for simplicity
$\beta_{K_1}=\beta_{K_{1A}}=\beta_{K_{1B}}=\beta_{K^*_2}$~\cite{ISGW2}
and use $f_{K_1(1270)}=175$~MeV~extracted from the measured
$\tau\to K_1(1270)\nu_\tau$ decays \cite{Cheng:DAP} to fix
$\beta_{K_1}$ to be $0.2979$~GeV.

\begin{table}[t!]
\caption{\label{tab:input} The input parameters $m_q$ and $\beta$
(in units of GeV) in the Gaussian-type wave function
(\ref{eq:wavefn}).}
\begin{ruledtabular}
\begin{tabular}{cccccc}
          $m_u$
          & $m_s$
          & $m_b$
          & $\beta_B$
          & $\beta_{K^*}$
          & $\beta_{K_1,K^*_2}$
          \\
\hline $0.23$
          & $0.45$
          & $4.4$
          & $0.5233$
          & $0.2846$
          & $0.2979$
\end{tabular}
\end{ruledtabular}
\end{table}

\begin{table}[b]
\caption{Tensor form factors of $B\to K^*,\,K_1,\,K^*_2$
transitions obtained in the covariant light-front model are fitted
to the 3-parameter form Eq. (\ref{eq:FFpara}) except for the form
factors $Y_{B3}$ and $U_2$ denoted by $^{*}$ for which the fit
formula Eq. (\ref{eq:FFpara1}) is used. All the form factors are
dimensionless.}
 \label{tab:FF}
\begin{ruledtabular}
\begin{tabular}{| c c c c c || c c c c c |}
~~~$F$~~~~~
    & $F(0)$~~~~~
    & $F(q^2_{\rm max})$~~~~
    &$a$~~~~~
    & $b$~~~~~~
& ~~~ $F$~~~~~
    & $F(0)$~~~~~
    & $F(q^2_{\rm max})$~~~~~
    & $a$~~~~~
    & $b$~~~~~~
 \\
    \hline
$T_1$
    & $0.24$
    & $1.00$
    & 1.73
    & 0.90
&$Y_{A1}$
    & 0.11
    & 0.15
    & 0.68
    & 0.35 \\
$T_2$
    & $0.24$
    & $0.59$
    & 0.92
    & 0.07
&$Y_{A2}$
    & 0.11
    & 0.06
    & $-0.91$
    & 0.79
    \\
$T_3$
    & 0.17
    & 0.79
    & 1.72
    & 0.84
&$Y_{A3}$
    & 0.19
    & 0.34
    & 1.02
    & 0.35
    \\
$Y_{B1}$
    & 0.13
    & 0.33
    & 1.94
    & 1.53
&$U_1$
    & 0.19
    & 0.45
    & 2.22
    & 2.13
    \\
 $Y_{B_2}$
    & 0.13
    & 0.21
    & 0.83
    & 0.25
&$U_2$
    & $0.19^*$
    & $0.32^*$
    & $1.77^*$
    & $4.32^*$
    \\
 $Y_{B3}$
    & $-0.07^*$
    & $-0.24^*$
    & $1.93^*$
    & $2.33^*$
&$U_3$
    & 0.16
    & 0.37
    & 2.19
    & 1.80
    \\
\end{tabular}
\end{ruledtabular}
\end{table}

As in \cite{Jaus99,CCH}, because of the condition $q^+=0$ we have
imposed during the course of calculation, form factors are known
only for spacelike momentum transfer $q^2=-q^2_\bot\leq 0$,
whereas only the timelike form factors are relevant for the
physical decay processes. It has been proposed in \cite{Jaus96} to
recast the form factors as explicit functions of $q^2$ in the
spacelike region and then analytically continue them to the
timelike region. It has been shown recently that, within a
specific model, form factors obtained directly from the timelike
region (with $q^+>0$) are identical to the ones obtained by the
analytic continuation from the spacelike region~\cite{BCJ03}.

In principle, form factors at $q^2>0$ can be evaluated directly in
the frame where the momentum transfer is purely longitudinal,
i.e., $q_\bot=0$, so that $q^2=q^+q^-$ covers the entire range of
momentum transfer \cite{Cheng97}. The price one has to pay is
that, besides the conventional valence-quark contribution, one
must also consider the non-valence configuration (or the so-called
$Z$-graph) arising from quark-pair creation from the vacuum.
However, a reliable way of estimating the $Z$-graph contribution
is still lacking unless one works in a specific model, for
example, the one advocated in \cite{BCJ03}. Fortunately, this
additional non-valence contribution vanishes in the frame where
the momentum transfer is purely transverse i.e., $q^+=0$.

\begin{table}[b]
\caption{Tensor form factor $T_1$ at $q^2=0$ in this work and in
various other models. }
 \label{tab:T1s}
\begin{ruledtabular}
\begin{tabular}{ll|ll}
 Ref.
      &$T_1(0)$
      &Ref.
      &$T_1(0)$
      \\
      \hline
 This work~~~~~
      & 0.24
      & LCSR~\cite{ABS94}
      & $0.32\pm0.05$
      \\
 QM \cite{MS}
      & $0.37\pm0.09,~0.39$
      & LCSR~\cite{Narison94}
      & $0.31\pm0.04$
      \\
 LFQM~\cite{ODonnell95,GHLZ01}
      & 0.32\footnotemark[1]
      & LCSR~\cite{BB98}
      & $0.38\pm0.06$
      \\
 Lattice~\cite{BHS93}
      & $0.20\pm0.02\pm0.06$~~~~~
      & LCSR~\cite{Safir01}
      & $0.32\pm0.06$
      \\
 Lattice~\cite{UKQCD98}
      & $0.32^{+0.04}_{-0.02}$
      & PQCD \cite{CG}
      & 0.315\footnotemark[2] (0.294)\footnotemark[3]
      \\
 Lattice~\cite{Becirevic03}
      & $0.25\pm0.05\pm0.02$ & &
      \\
\end{tabular}
\end{ruledtabular}
 \footnotetext[1] {For $\beta_{K^*}=0.32$ GeV.}
 \footnotetext[2] {For $\beta_{B}=0.40$ GeV.}
  \footnotetext[3] {For $\beta_{B}=0.42$ GeV.}
\end{table}

To proceed we find that except for the form factors $Y_{B3}$ and
$U_2$, the momentum dependence of the form factors
$T_i,\,Y_{Ai,Bi},\,U_i$ in the spacelike region can be well
parameterized and reproduced in the three-parameter form:
 \be \label{eq:FFpara}
 F(q^2)=\,{F(0)\over 1-a(q^2/m_{B}^2)+b(q^2/m_{B}^2)^2}\,.
 \en
The parameters $a$, $b$ and $F(0)$ are first determined in the
spacelike region. We then employ this parametrization to determine
the physical form factors at $q^2\geq 0$. In practice, the
parameters $a,b$ and $F(0)$ are obtained by performing a
3-parameter fit to the form factors in the range $-20\,{\rm
GeV}^2\leq q^2\leq 0$. The obtained $a$ and $b$ parameters are in
most cases are not far from unity as expected. However, the
parameter $b$ for $Y_{B3}$ and $U_2$ is rather sensitive to the
chosen range for $q^2$ and can be as large as 6.6 and 8.8,
respectively. To overcome this difficulty, we will fit
$Y_{B3}(q^2)$ and $U_2(q^2)$ to the form
 \be \label{eq:FFpara1}
 F(q^2)=\,{F(0)\over (1-q^2/m_{B}^2)[1-a(q^2/m_{B}^2)+b(q^2/m_{B}^2)^2]}
 \en
and achieve a substantial improvement. Note that for the case of
$U_2(q^2)$, it is fitted to a smaller range of $-12\,{\rm
GeV}^2\leq q^2\leq 0$.

\begin{figure}[t]
\centerline{
            {\epsfxsize3 in \epsffile{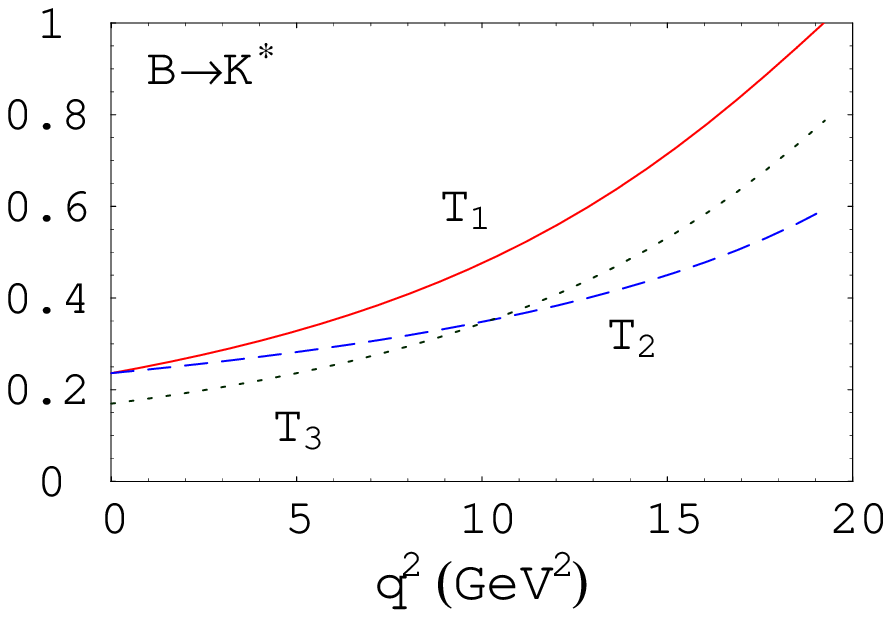}}
            {\epsfxsize3 in \epsffile{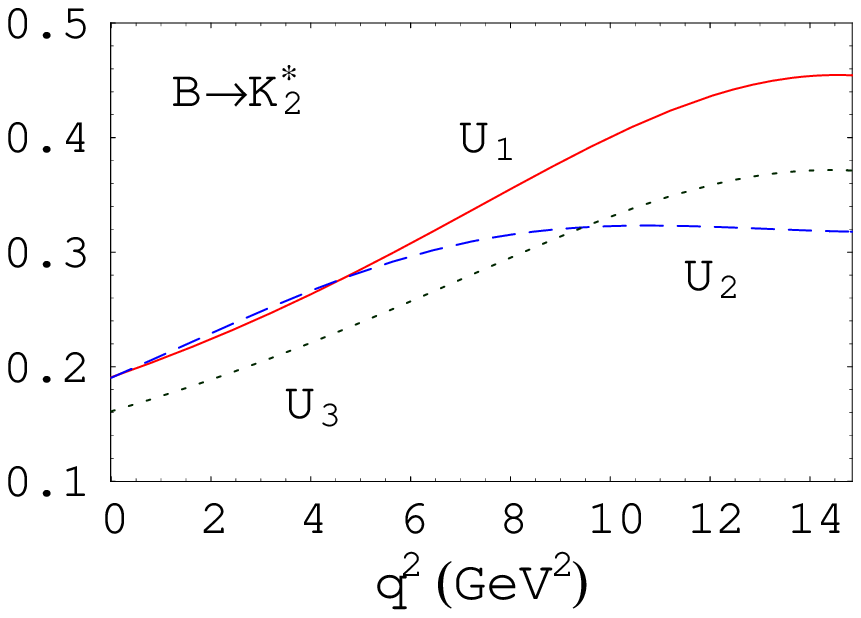}}}
\centerline{
            {\epsfxsize3 in \epsffile{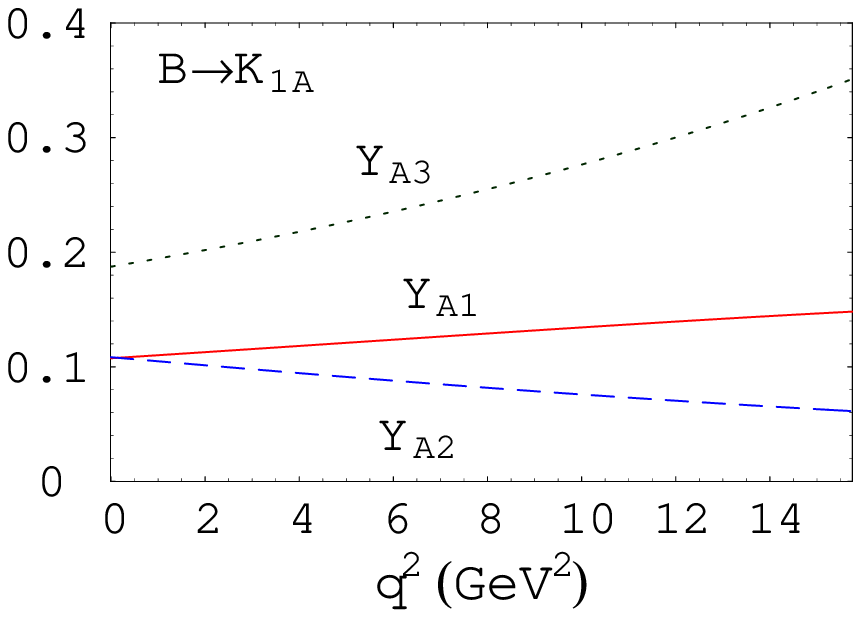}}
            {\epsfxsize3 in \epsffile{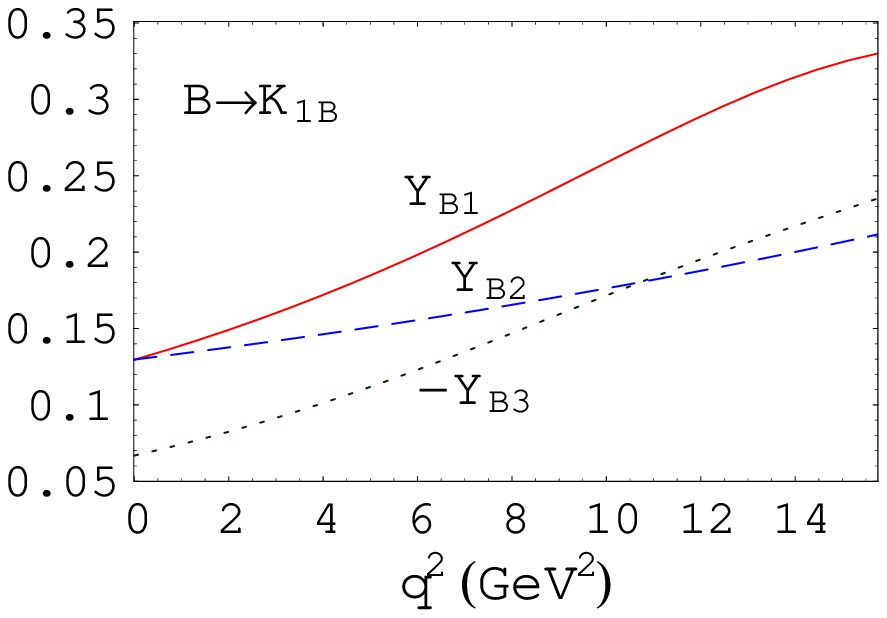}}}
\caption{Tensor form factors $T_{i}(q^2)$, $Y_{Ai,Bi}(q^2)$ and
$U_i(q^2)$ for $B\to K^*$, $B\to K_1$ and $B\to K_2^*$
transitions, respectively.} \label{fig:FF} 
\end{figure}

The $B\to K^*,\,K_1,\,K^*_2$ transition tensor form factors and
their $q^2$-dependence are displayed in Table~\ref{tab:FF} and
depicted in Fig.~\ref{fig:FF}. The result of $T_1(0)$ is then
compared with other model calculations in Table~\ref{tab:T1s}. It
should be stressed that our $T_1(0)$ is smaller than that obtained
from the quark model (QM) \cite{MS}, the conventional light-front
quark model (LFQM)~\cite{ODonnell95,GHLZ01}\footnote{This is due
to the presence of additional terms proportional to
$(1-x)m_q^2+p'^2_\bot$ in the expression of the form factor
$T_1(0)$ in the conventional light-front model [see in Eq.
(\ref{eq:T1ODonnell})]. However, it is shown in Appendix that this
tensor form factor does not have the correct heavy quark limit
behavior. In the heavy quark limit, heavy quark spin symmetry
allows one to relate the tensor form factors $T_i(q^2)$ to the
vector and axial-vector $B\to V$ transition form factors (see
Appendix). As shown in~\cite{CCH}, the latter in the covariant
light-front model are numerically smaller than the other model
results. Therefore, the fact that $T_1(0)$ is smaller in the
covariant LF model (see Table~\ref{tab:FF}) is consistent with
previous form factor calculations in~\cite{CCH}.}, light-cone sum
rules (LCSR)~\cite{ABS94,Narison94,BB98,Safir01} and the
perturbative QCD approach (PQCD) \cite{CG} but is close to two of
the lattice calculations~\cite{BHS93,Becirevic03}.

The effective parameter $a_{11}(K^*\gamma)$ has been calculated in
the framework of QCD factorization to be
$|a_{11}(K^*\gamma)|^2=0.165^{+0.018}_{-0.017}$~\cite{QCDfacBFS}
at $\mu=\hat m_b$ and
$a_{11}(K^*\gamma)=-0.4072-0.0256i$~\cite{QCDfacBB} at $\mu=m_b$
(see also Table 3 of \cite{Ali}). These effective parameters are
larger than the Wilson coefficient $c_{11}$ of order $-0.32$ at
$\mu=m_b$. For $B\to K_1\gamma$ and $K^*_2\gamma$ decays, we shall
employ $a_{11}=c_{11}$ as NLO QCD corrections there have not been
calculated. In Eq.~(\ref{eq:Br}), we take $m_b(\hat m_b)=4.4$~GeV
for $B\to K^*\gamma$ and $m_b(m_b)=4.2$~GeV for $B\to K_1\gamma$
and $K_2^*\gamma$ decays.

\begin{table}[t]
\caption{Branching ratios for the radiative decays $B\to
K^*\gamma,\,K_1(1270)\gamma,\,K_1(1400)\gamma,\,K^*_2(1430)\gamma$
(in units of $10^{-5}$) in the covariant light-front model and in
other models. Experimental data are summarized in Sec. I and only
the averages for $B\to K^*\gamma$ and $B\to K_2^*\gamma$ are
quoted in the table. Experimental limits on $B\to K_1\gamma$ are
taken from \cite{Belle02}.}
 \label{tab:Br}
\begin{ruledtabular}
\begin{tabular}{ l l l l l }
~~~~~~~~~~~~~~~~~~~~
    & $B\to K^*\gamma$~~~~~
    & $B\to K_1(1270)\gamma$~~~~
    & $B\to K_1(1400)\gamma$~~~~~
    & $B\to K^*_2(1430)\gamma$~~~~~~
 \\
    \hline
Expt
    & $4.17\pm0.19$
    & $<9.9$
    & $<5.0$
    & $1.33\pm0.20$
\\
This work
    & $3.27\pm0.74\footnotemark[1]$
    & $0.02\pm0.02\footnotemark[2]$
    & $0.80\pm0.12$\footnotemark[2]
    & $1.48\pm0.30$
\\
    &
    & $(0.04\pm0.03)\footnotemark[2]$
    & $(0.77\pm0.11)$\footnotemark[2]
    &
\\
    &
    & $0.77\pm0.11$\footnotemark[3]
    & $0.08\pm0.04$\footnotemark[3]
    &
\\
    &
    & $(0.84\pm0.12)$\footnotemark[3]
    & $(0.003\pm0.006)$\footnotemark[3]
    &
    \\
Lattice~\cite{Becirevic03}
    & $3.54\pm 1.57\footnotemark[1]$
    &
    &
    &
    \\
RQM~\cite{EFGT01}
    & $4.5\pm1.5$
    & $0.45\pm0.15$
    & $0.78\pm0.18$
    & $1.7\pm0.6$
    \\
LFQM~\cite{GHLZ01}
    & $5.81\pm1.32\footnotemark[1]$
    &
    &
    &
    \\
LCSR~\cite{Safir01}
    & $5.81\pm2.27\footnotemark[1]$
    & $0.67\pm0.27\footnotemark[4]$
    & $0.30\pm0.13\footnotemark[4]$
    & $1.67\pm0.67\footnotemark[4]$
    \\
AP~\cite{Ali}
    & $6.8\pm 2.6$
    &
    &
    &
    \\
BB~\cite{QCDfacBB,BB03}
    & $7.4^{+2.6}_{-2.4}\footnotemark[5]$
    &
    &
    &
    \\
BFS~\cite{QCDfacBFS}
    & $7.9^{+3.5}_{-3.0}$
    &
    &
    &
    \\
HQET~\cite{VO96}
    & $9.99\pm3.81\footnotemark[6]$
    & $1.44\pm0.53\footnotemark[4]$
    & $0.70\pm0.30\footnotemark[4]$
    & $2.07\pm0.97\footnotemark[4]$
    \\
\end{tabular}
\end{ruledtabular}
 \footnotetext[1] {Use of
 $|a_{11}(K^*\gamma)|^2=0.165\pm0.018$~\cite{QCDfacBFS} and
 Eq.~(\ref{eq:Br}) has been made.}
 \footnotetext[2]{For the $K_1(1270)$--$K_1(1400)$ mixing angle $\theta=-58^\circ(-37^\circ)$.}
 \footnotetext[3]{For the $K_1(1270)$--$K_1(1400)$ mixing angle $\theta=+58^\circ(+37^\circ)$.}
 \footnotetext[4]{Use has been made of ${\mathcal B}(b\to s\gamma)=3.34\times
 10^{-4}$~\cite{Nishida03}.}
 \footnotetext[5]{The central value and errors are taken from the complete NLO
 result for the neutral mode \cite{BB03}. }
 \footnotetext[6]{The original results are scaled up by a factor of
 $|a_{11}(K^*\gamma)/c_{11}|^2=1.78$\,.}

\end{table}

In Table~\ref{tab:Br}, we summarize the calculated branching
ratios for the radiative decays $B\to K^*\gamma$,
$K_1(1270)\gamma$, $K_1(1400)\gamma$, $K^*_2(1430)\gamma$ in the
covariant light-front model. For comparison we also quote
experimental results and some other theoretical calculations. For
our results and results in  LFQM~\cite{GHLZ01},
lattice~\cite{Becirevic03} and LCSR~\cite{Safir01}, we use
$|a_{11}(K^*\gamma)|^2=0.165\pm0.018$~\cite{QCDfacBFS}. The
theoretical errors in  ${\mathcal B}(B\to K^*\gamma)$ arise from
$|a_{11}(K^*\gamma)|^2$ and $T_1(0)$. Note that we have assigned a
10\% estimated uncertainty for our $T_1(0)$ and that from
LFQM~\cite{GHLZ01}. For $B\to K^*\gamma$ rates from the
relativistic quark model (RQM)~\cite{EFGT01} and heavy quark
effective theory (HQET)~\cite{VO96}, we have scaled up their
results by a factor of $|a_{11}(K^*\gamma)/c_{11}|^2=1.78$.
Calculations in LCSR~\cite{Safir01} and HQET~\cite{VO96} are often
expressed in terms of $R\equiv{\mathcal B}(B\to
K^{**}\gamma)/{\mathcal B}(b\to s\gamma)$ with $K^{**}$ denoting
$K_1$ or $K_2^*$. Therefore, the branching ratio of $B\to
K^{**}\gamma$ is obtained by multiplying $R$ with ${\mathcal
B}(b\to s\gamma)=3.34\times 10^{-4}$~\cite{Nishida03}. For $B\to
K^*_2\gamma$, the error in our predicted rate shown in
Table~\ref{tab:Br} comes from a 10\% estimated uncertainty in
$U_1(0)$.

As stressed in \cite{QCDfacBFS,QCDfacBB,Ali}, the NLO correction
yields an enhancement of the $B\to K^*\gamma$ rate that can be as
large as 80\%. Consequently, the prediction in most of the
existing models becomes too large as the measured branching ratio
is already saturated even before the NLO correction is taken into
account. Our prediction of $\B(B\to K^*\gamma)=(3.27\pm0.74)\times
10^{-5}$ due to short-distance $b\to s\gamma$ contributions agrees
with experiment (see Table ~\ref{tab:Br}).  It is generally
believed that long-distance contributions to $B\to K^*\gamma$ is
small and not more than 5\% (see e.g. \cite{Grinstein,KRSW} and
references therein).

To compute $B\to K_1\gamma$ rates we need to know the
$K_1(1270)$--$K_1(1400)$ mixing angle as defined in Eq.
(\ref{eq:K1mixing}). From the experimental information on masses
and the partial rates of $K_1(1270)$ and $K_1(1400)$, Suzuki found
two possible solutions with a two-fold ambiguity, $|\theta|\approx
33^\circ$ and $57^\circ$~\cite{Suzuki}. A similar constraint
$35^\circ\lsim |\theta|\lsim 55^\circ$ is obtained in
\cite{Goldman} based solely on two parameters: the mass difference
of the $a_1$ and $b_1$ mesons and the ratio of the constituent
quark masses. An analysis of $\tau\to K_1(1270)\nu_\tau$ and
$K_1(1400)\nu_\tau$ decays also yields the mixing angle to be
$\approx 37^\circ$ or $58^\circ$ with a two-fold ambiguity
\cite{Cheng:DAP}. It has been shown in \cite{Cheng:DAP} that the
study of hadronic decays $D\to K_1(1270)\pi,~K_1(1400)\pi$ decays
favors the solution $\theta\approx -58^\circ$. However, this is
subject to many uncertainties such as the unknown $D\to K_{1A,1B}$
transition form factors and the decay constants of $K_1(1270)$ and
$K_1(1400)$.

The physical $B\to K_1(1270)$ and $B\to K_1(1400)$ tensor form
factors have the expressions
 \be \label{eq:Ymixing}
 Y_i^{B\to K_1(1270)}(q^2) &=&
 Y_{Ai}(q^2)\sin\theta+Y_{Bi}(q^2)\cos\theta \non \\
 Y_i^{B\to K_1(1400)}(q^2) &=&
 Y_{Ai}(q^2)\cos\theta-Y_{Bi}(q^2)\sin\theta.
 \en
Since the form factors $Y_{A1}(0)$ and $Y_{B1}(0)$ are similar
(see Table \ref{tab:FF}) and since the $K_1(1270)$--$K_1(1400)$
mixing angle is large, it is obvious from Eqs. (\ref{eq:Ymixing})
and (\ref{eq:Br}) that one of $B\to K_1\gamma$ decays is strongly
suppressed owing to a large cancellation between the $Y_{A1}(0)$
and $Y_{B1}(0)$ terms. In Table \ref{tab:Br}, branching ratios of
$B\to K_1\gamma$ are calculated using two different sets of the
$K_1(1270)$--$K_1(1400)$ mixing angles $\theta=\pm 58^\circ,\pm
37^\circ$. \footnote{Note that by using input parameters in
Table~\ref{tab:input} and with four different values of $\theta$,
the decay constant $|f_{K_1(1270)}|$ is still within the
experimental range ($175\pm19$)~MeV.} Errors in the rates
displayed in Table \ref{tab:Br} stem from 10\% estimated
uncertainties in $Y_{A1,B1}(0)$. Therefore, the ratio of $B\to
K_1(1270)\gamma$ and $K_1(1400)\gamma$ rates is very sensitive to
the mixing angle. For example for $\theta=\pm 58^\circ$ we have
 \be
 \frac{{\mathcal B}(B\to K_1(1270)\gamma)}{{\mathcal
 B}(B\to K_1(1400)\gamma)}=\cases{10.1\pm6.2 &
 for~$\theta=+58^\circ$, \cr
 0.02\pm0.02  & for~$\theta=-58^\circ$. }
 \en
Evidently, experimental measurement of the above ratio of
branching fractions can be used to fix the sign of the mixing
angle, and it should be much more clean than the method based on
hadronic $D$ decays~\cite{Cheng:DAP}.

It is worth emphasizing that all other models predict comparable
$K_1(1270)\gamma$ and $K_1(1400)\gamma$ rates (see Table
\ref{tab:Br}). In \cite{Safir01,VO96} tensor form factors $Y_i$
are evaluated directly for the physical $B\to K_1(1270)$ and $B\to
K_1(1400)$ transitions, while $B\to K_1^{1/2}$ and $B\to
K_1^{3/2}$ transition form factors ($K_1^{1/2}$ and $K_1^{3/2}$
being the $P_1^{1/2}$ and $P_1^{3/2}$ states of $K_1$,
respectively) are evaluated first in~\cite{EFGT01} and then
related to the physical transitions. Hence, measurements of $B\to
K_1\gamma$ decays can be utilized to distinguish the covariant
light-front model from others.

For $B\to K_2^*\gamma$ decays, the calculated branching ratio of
$(1.48\pm0.30)\times 10^{-5}$ is in a good agreement with the
world average of $(1.33\pm0.20)\times 10^{-5}$. Since the above
prediction is for $a_{11}=c_{11}$, this seems to imply that NLO
corrections to $B\to K_2^*\gamma$ is not as important and dramatic
as in the case of $B\to K^*\gamma$.

\section{Conclusion}
 Exclusive radiative $B$ decays, $B\to K^*\gamma$, $K_1(1270)\gamma$,
$K_1(1400)\gamma$ and $K_2^*(1430)\gamma$, are studied in the
framework of a covariant light-front quark model. Our main
conclusions are as follows.
 \begin{enumerate}
 \item
The tensor form factor $T_1(q^2)$ at $q^2=0$, which is relevant to
the decay $B\to K^*\gamma$, is found to be 0.24\,. This is much
smaller than what expected from the conventional light-front model
or light-cone sum rules but is in a good agreement with a recent
lattice result \cite{Becirevic03}. In the heavy quark limit, the
tensor form factors can be related to the vector and axial-vector
form factors. Contrary to the conventional light-front model, it
is found that the expression of $T_1(q^2)$ in the covariant
light-front model has the correct heavy quark limit behavior.
 \item
Taking into account the next-to-leading order hard vertex and hard
spectator corrections, the predicted branching ratio $\B(B\to
K^*\gamma)=(3.27\pm 0.74)\times 10^{-5}$ agrees with experiment,
whereas most of the existing models predict too large decay rates
of $B\to K^*\gamma$ compared to the data.
 \item
The decay rates of $B\to K_1(1270)\gamma$ and $B\to
K_1(1400)\gamma$ are very sensitive to the
$K_1(1270)$--$K_1(1400)$ mixing angle and hence a measurement of
their relative strength will provide an excellent way for
determining the sign of the strange axial-vector meson mixing
angle. Contrary to the other models in which $K_1(1270)\gamma$ and
$K_1(1400)\gamma$ are predicted to be comparable, we found that,
depending on the sign of the mixing angle, one of them is strongly
suppressed owing to a large cancellation between two different
form factor terms. Hence experimental measurements of the ratio of
branching fractions will enable us to discriminate between
different models.
 \item
The predicted branching ratio of $B\to K_2^*\gamma$ is in a good
agreement with experiment and this may imply that NLO corrections
to $B\to K_2^*\gamma$ is not as important and dramatic as in the
case of $B\to K^*\gamma$.

 \end{enumerate}

\vskip 2.5cm \acknowledgments We are grateful to Chuang-Hung Chen
for valuable discussions. This research was supported in part by
the National Science Council of R.O.C. under Grant Nos.
NSC92-2112-M-001-016 and NSC92-2811-M-001-054.

\newpage
\appendix
\section{Heavy quark limit of the form factor $T_1(0)$}
In the heavy quark limit the tensor form factors $T_i(q^2)$ for
$B\to K^*$ transition can be related to vector and axial-vector
$B\to K^*$ form factors defined by
 \be
  \la K^*(P^\pp,\vp^\pp)|V_\mu|B(P^\prime)\ra
          &=&\epsilon_{\mu\nu\alpha \beta}\,\vp^{\pp*\nu}P^\alpha q^\beta\, g({q^2}),
\non\\
\la K^*(P^\pp,\vp^\pp)|A_\mu|B(P^\prime)\ra
          &=&-i\left\{\varepsilon_\mu^{\pp*} f({q^2})
              +\vp^{*\pp}\cdot P \left[P_\mu a_+({q^2})+q_\mu
              a_-({q^2})\right]\right\}.
 \en
In the static limit of the $b$ quark, the static $b$-quark spinor
satisfies the equation of motion $\gamma_0 b=b$. Heavy quark spin
symmetry implies the relations \cite{IW89}
 \be
 \la \ov K^*|\bar s\gamma_i b|\ov B\ra &=& \la \ov K^* |\bar
 si\sigma_{0i}b|\ov B\ra, \non \\
 \la \ov K^*|\bar s\gamma_i\gamma_5 b|\ov B\ra &=& -\la \ov K^* |\bar
 si\sigma_{0i}\gamma_5b|\ov B\ra.
 \en
This gives the form factor relation (see e.g. \cite{ODonnell93}
for other form-factor relations)
 \be
 T_1(q^2)=-{1\over 2}(m_B-\omega\, m_{K^*})g(q^2)-{1\over 4m_B}f(q^2),
 \en
where
 \be
 \omega={m_B^2+m_{K^*}^2-q^2\over 2m_B m_{K^*}}.
 \en
Then in the heavy quark limit
 \be \label{eq:HQS}
 T_1(q^2)=-{1\over 4}m_b\,g(q^2)-{1\over 4m_b}f(q^2)
 \en
for $|q^2|\ll m_B^2$.

From Eq. (\ref{eq:T1}) we find that in the heavy quark limit
 \be \label{eq:T1HQlimit}
 T_1(0)\to  {1\over 32\pi^2}\int dx\, d^2p'_\bot {xm_bm_q\,\varphi^\pp\varphi'
 \over \sqrt{{\cal A}'^2+p'^2_\bot}\sqrt{{\cal A}^{\pp
 2}+p'^2_\bot}},
 \en
where use of $x\to 0$ has been made. It follows from Eq. (B4) of
\cite{CCH} that
 \be
 g(0)\to -{1\over 16\pi^2}\int dx\, d^2p'_\bot {\varphi^\pp\varphi'
 \over \sqrt{{\cal A}'^2+p'^2_\bot}\sqrt{{\cal A}^{\pp
 2}+p'^2_\bot}}\,\Big[x^2m_b+xm_q+{p'^2_\bot+m_q^2-x^2m_b^2\over m_b}\Big],
 \en
and
 \be
  f(0) &\to& {1\over 32\pi^2}\int dx\, d^2p'_\bot {\varphi^\pp\varphi'
 \over \sqrt{{\cal A}'^2+p'^2_\bot}\sqrt{{\cal A}^{\pp
 2}+p'^2_\bot}} \non \\
 &\times& \Big[2xm_b^2(xm_b-m_q)+2\,{p'^2_\bot+m_q^2-x^2m_b^2\over m_b} \Big].
 \en
Hence,
 \be \label{eq:g&f}
 -{1\over 4}m_b\,g(0)-{1\over 4m_b}f(0) \to {1\over 32\pi^2}\int dx\, d^2p'_\bot
 {xm_b m_q\,\varphi^\pp\varphi'
 \over \sqrt{{\cal A}'^2+p'^2_\bot}\sqrt{{\cal A}^{\pp
 2}+p'^2_\bot}}.
 \en
By comparing (\ref{eq:g&f}) with (\ref{eq:T1HQlimit}) we see that
$T_1$ has the correct heavy quark limit behavior. It should be
stressed that the zero mode contribution to the form factor
$f(q^2)$ vanishes in the heavy quark limit. In the conventional
light-front model \cite{ODonnell95,GHLZ01}, the heavy quark limit
of $T_1(0)$ contains an additional term $m_q^2+p'^2_\bot$ in the
numerator of Eq. (\ref{eq:T1HQlimit}) [see Eq.
(\ref{eq:T1ODonnell})] and hence it does not respect
(\ref{eq:HQS}) in the heavy quark limit.

\end{document}